# New technologies and AI: envisioning future directions for UNSCR 1540

Clara Punzi[a,b]

[a] Scuola Normale Superiore, P.za dei Cavalieri, 7, 56126, Pisa, Italy
[b] KDD Lab, ISTI-CNR, Via G. Moruzzi 1, 56124 Pisa, Italy

clara.punzi@sns.it

## Introduction

Amidst a time characterized by unprecedented technological progress, the integration of Artificial Intelligence (AI) in the military domain has emerged as a powerful factor capable of reshaping several facets of global security and altering the dynamics of geopolitics and warfare[1]. Indeed, AI has the potential to protect lives and prevent various threats, such as physical and cybersecurity attacks. However, it may also be utilized to enhance the destructive capabilities of systems, as exemplified by kamikaze drones[2] and killer robots[3].

When the United Nations Security Council (UNSC) first adopted Resolution 1540 (hereinafter UNSCR 1540) in 2004, the aim was to create a pivotal framework to address and diminish the threat presented by the proliferation of weapons of mass destruction (WMDs), specifically nuclear, chemical, and biological weapons, along with their means of delivery. At that time, the threat posed by a possible interaction with AI was not contemplated. Nevertheless, the growing capability and ubiquity of AI have brought a new level of complexity to the implementation and outcomes of this resolution, resulting in the acknowledgment, in subsequent resolutions, of potential adverse effects posed by the use of AI in the context of implementation of UNSCR 1540. Recently, these concerns have been further highlighted by the UNSCR during the first debate on AI, peace and security, where the United Nations Secretary-General expressed deep alarm regarding the interaction between AI and nuclear weapons, biotechnology, neurotechnology, and robotics[4]. Despite this emergent awareness, there is at the moment a lack of international, multilateral agreements or any existing governance framework that covers the use of AI in the military, as also highlighted during the first conference on Responsible AI in the Military Domain[5].

This work delves into the concerns surrounding the intersection of AI and UNSCR 1540, exploring the potential risks and ethical considerations that characterize the evolving landscape of global security. We critically examine the implications of AI on the goals and efficacy of UNSCR 1540 in safeguarding the world from the threats posed by WMDs proliferation. More precisely, we argue that AI can both amplify pre-existing risks associated with WMDs and also generate novel hazards. In both cases, the use of AI would create or exacerbate existing threats to international peace and security. As a result, we contend that there is a need to broaden the scope of UNSCR 1540 to specifically encompass the impact of AI technologies on the development, distribution, and proliferation of WMDs, with the aim of averting catastrophic consequences.

---

[1] https://www.ceps.eu/why-the-eu-must-now-tackle-the-risks-posed-by-military-ai/

[2] https://futurium.ec.europa.eu/en/european-ai-alliance/blog/challenges-governing-ai-military-purposes-and-spill-over-effects-ai-act?language=en

[3] The term "Killer robots" is an informal expression referring to lethal autonomous weapons systems (LAWS) capable of independently identifying, targeting, and attacking humans or objects without direct human involvement. See also https://www.stopkillerrobots.org/

[4] https://www.un.org/sg/en/content/sg/speeches/2023-07-18/secretary-generals-remarks-the-security-council-artificial-intelligence

[5] https://reaim2023.org/

## AI technologies within 1540 resolution

Prior to analyzing the various ways in which AI relates to the scope of UNSCR 1540, it is essential to assess how UNSCR 1540 and subsequent resolutions embrace the role of new technologies, especially AI. By doing so, in the next sections, we will be able to identify the current gaps in this legal framework.

As mentioned in the Introduction, a first relevant observation is that AI systems can be regarded as "means of delivery"[6] according to UNSCR 1540 if they are autonomous and have been designed for the scope of delivering WMDs. If this is the case, then all provisions of UNSCR 1540 and following resolutions apply. Successively, the UNSC also affirms to be gravely concerned by the risk of state[7] and non-State[8] actors using the rapid advances in science and technology to ends related to the proliferation of WMDs. Furthermore, UNSCR 1540 and subsequent resolutions emphasize that the progress of innovative technology for peaceful purposes should not be hindered, but the pursuit of peaceful applications should not be exploited as a disguise for the spread of WMDs[9].

This last observation raises concerns about *dual-use AI technologies,* that is, AI applications that can be employed for positive, constructive purposes as well as for activities that may pose risks or have negative consequences on human lives. Such a duality arises from the fact that the impact of AI technologies often depends on how they are deployed and for what purpose. For example, generative AI models are considered dual-use: while positive applications exist, such as content creation, translation services, and creative projects, they can also be misused for generating deceptive content, deepfakes, or spreading misinformation. Similarly, AI applications in fields like cybersecurity, autonomous vehicles, and healthcare may have dual-use characteristics: while these technologies can enhance security[10], transportation efficiency, and medical diagnostics, potential malicious uses, such as cyber attacks, weaponization of autonomous systems, or privacy violations, are also possible. The dual-use nature of many AI technologies highlights the significance of ethical considerations, responsible development, and appropriate legislation to guide and regulate their design, implementation, and deployment. This is crucial in order to maximize the potential advantages while minimizing the associated risks and negative consequences. Interestingly, dual-use technologies seem to be excluded from UNSCR 1540, since all provisions refer to applications with the specific, hence not dual, purpose of enhancing the proliferation of WMDs. This constitutes an important gap in the regulatory framework.

---

[6] UNSCR 1540 defines "means of delivery" as «missiles, rockets and other *unmanned systems* capable of delivering nuclear, chemical, or biological weapons, that are specially designed for such use.»
[7] S/RES/2325 (2016), para. 7 and S/RES/2663 (2022), para. 14.
[8] S/RES/2325 (2016), para. 8 and S/RES/2663 (2022), para. 15.
[9] S/RES/1540 (2004), S/RES/1810 (2008) and S/RES/2325 (2016)
[10] Jada, I., & Mayayise, T. O. (2023). The impact of artificial intelligence on organisational cyber security: An outcome of a systematic literature review. In Data and Information Management (p. 100063). Elsevier BV. https://doi.org/10.1016/j.dim.2023.100063

However, it is important to acknowledge that other legislative structures are being developed, at national and regional level, that may add elements to the governance of AI systems with some application in the domain of WMDs. For instance, the European Union proposal for an Artificial Intelligence Act, while excluding the military domain from its scope, states that dual-use AI systems are required to comply with its provisions for high-risk AI[11], which include a number of ethical and legal standards that should guarantee respect for fundamental human rights.

## AI technologies as means of delivery of WMDs

In the perspective of limiting the harms provoked by the use of WMDs, AI technologies could be employed to improve targeting accuracy, reduce unintended damage, and accelerate decision-making processes. At the same time, such AI-controlled WMDs could result in greater lethality and destruction. Indeed, it is widely acknowledged in the AI community that the use of AI in critical domains poses significant risks due to the potential for costly mistakes, which may happen as a result of prediction errors or unintended discriminations that may arise from hidden flaws in the machine learning process. The military domain here under analysis is not exempt from such risks[12]. For instance, if a state or non-state actor has the capability to gather data, such as biometric data, about a subordinate population through a digital surveillance system[13], they could potentially use this information to launch a military attack against the entire population without differentiating between military and civilian targets. This could lead to catastrophic consequences, not only from a humanitarian and ethical standpoint but also in terms of international law.

The detrimental impacts of AI applications as means of delivery of WMDs become particularly dangerous in the case of autonomous decision-making systems. In other words, when the algorithm is granted the power to select and attack certain targets with WMDs. In this scenario, the absence of human control highlights and exacerbates several concerns from technical, ethical, and legal perspectives. From a technical perspective, there is a potential for unintended negative consequences to occur when the AI responds to real-world situations that it has not been trained for or lacks confidence in handling. These consequences may include harm to civilians and damage to infrastructure. Secondly, it is crucial to acknowledge the enormous difficulty of constructing an autonomous AI system that consistently upholds ethical principles and human rights in all possible scenarios. Indeed, despite the considerable efforts made by the AI community to create "ethical machines", these endeavors usually necessitate human supervision and are full of weaknesses. Finally, questions regarding the accountability and legal obligations associated with the decisions taken also arise. In general, both deployment and decision-making processes should be guided by the principles of proportionality, distinction, and adherence to international humanitarian law. However, the development and use of AI-controlled WMDs inherently give rise to profound ethical questions, the foremost of which is whether it is worthwhile to use AI to augment the means of delivery of WMDs.

---

[11] https://futurium.ec.europa.eu/en/european-ai-alliance/blog/challenges-governing-ai-military-purposes-and-spill-over-effects-ai-act?language=en

[12] Schraagen, J. M. (2023). Responsible use of AI in military systems: prospects and challenges. In Ergonomics (Vol. 66, Issue 11, pp. 1719–1729). Informa UK Limited. https://doi.org/10.1080/00140139.2023.2278394

[13] This is the case, for instance, of the apartheid system entrenched by Israeli authorities over the Occupied Palestinian Territories (OPT), which largely employs face recognition technologies to segregate and control Palestinians in the OPT. See also https://www.amnesty.org/en/documents/mde15/6701/2023/en/

## AI technologies as WMD

Apart from serving as a means of delivery, certain technological advancements, and specifically AI, can potentially become WMD themselves, for instance by lowering the technical barriers to proliferation of traditional WMDs or by creating novel WMDs. In the first case, new technologies, such as computer numerical control[14], additives manufacturing[15], and synthetic biology[16], may facilitate the mass production of components of WMDs or may be used to more efficiently produce related materials[17].

Another manner in which AI connects with WMDs is via Generative AI (GAI), which refers to algorithms capable of generating seemingly novel content, including text, graphics, and audio, by matching user prompts with patterns learned during training. In this scenario, there is a concern that GAI could turn into a WMD itself rather than contributing to the development and proliferation of conventional WMDs. In fact, the inherent capability of generative AI to autonomously create, mimic, and manipulate information poses multifaceted challenges. Indeed, a ranking of top risks of 2023[18] included new technologies, and GAI in particular, as potential WMDs due to their significant disruptive impact on society. According to the authors of the report, GAI has the capacity to manipulate individuals and incite political instability. More precisely, the use of GAI, specifically through the creation and dissemination of fake, inaccurate, and deceptive material that appears convincing and authentic, can be utilized by demagogues and populists to manipulate society and undermine the integrity of information systems, potentially leading to the subversion of democracy. Additionally, autocrats can exploit GAI to establish digital surveillance and propaganda on social media platforms.

On a similar vein, the Global Network on Extremism and Technology[19] affirmed that GAI can facilitate extremist propaganda by desensitizing individuals. For instance, this can be accomplished by means of interactive social media, music and video games that represent extremist views and allow the player to identify with aggressive avatars. Another example involves the utilization of GAI to generate conspiratorial and violent digital content, which can also serve as material for political extremist propaganda.

---

[14] Computer numerical control is a technology that utilizes computerized systems to control and automate the operation of machine tools in manufacturing processes based on numerical instructions.
[15] Additive manufacturing, commonly known as 3D printing, is a technique that involves constructing objects by adding material layer by layer based on digital 3D models.
[16] Synthetic biology is an interdisciplinary field that encompasses the planning, assembly, and refinement of biological components and systems for innovative purposes, frequently with the objective of manufacturing synthetic organisms or augmenting preexisting biological capabilities.
[17] Stewart, I. J. (2018). Preventing WMD Proliferation: The Future of UNSCR 1540. In Preventing the Proliferation of WMDs (pp. 105–126). Springer International Publishing. https://doi.org/10.1007/978-3-319-72203-0_6
[18] https://www.eurasiagroup.net/issues/top-risks-2023
[19] https://gnet-research.org/2023/02/17/weapons-of-mass-disruption-artificial-intelligence-and-the-production-of-extremist-propaganda/

Global leaders in the development of GAI have also begun to shed light on the hazards that new technologies represent to the military realm. In late 2023, OpenAI, one of such players, released its "Preparedness Framework"[20]. This report outlines the processes that the company wants to implement in order to monitor, assess, predict, and safeguard against the potential dangers posed by increasingly powerful models. Notably, AI-assisted creation of WMDs is mentioned as one of the four risk categories to be tracked. Currently, it is awarded a low risk score, meaning that GAI models can provide individuals with information about the creation of WMDs that is similar to what can be obtained from conventional sources like search engines. Nevertheless, this assessment is prone to revision since the evaluation is expected to be consistently updated as the technology advances.

Other technologies can be exploited as WMDs in conjunction with GAI. This is the case of AI-assisted cybersecurity, the second risk category identified by OpenAI. More precisely, they refer to the use of GAI models to assist cyberattacks to disrupt confidentiality, integrity, and/or availability of computer systems. Currently, cybersecurity is awarded a medium risk score, which indicates that not only GAI models can meaningfully assist cyberattacks through non-programming tasks, but they can also increase the productivity of operators by a certain efficiency threshold on key cyber operation tasks.

## Conclusion

Throughout this work, we explored how new technologies, in particular AI, may impact the object and scope of implementation of UNSCR 1540. We reviewed how this resolution and subsequent ones address the issue and presented two possible interactions: AI-controlled WMDs and AI as WMD. Both cases raise questions about potential risks and ethical considerations with implications on global peace and security that are not fully acknowledged in an international regulatory framework. Consequently, we argue that it is necessary to expand the scope of UNSCR 1540 to explicitly include the influence of AI technology on the advancement, development, dissemination, and spread of WMDs in order to achieve effective and responsible global governance. Furthermore, we recommend the development of guidelines to ensure the effective implementation of such new provisions of Resolution 1540.

## Acknowledgment

I would like to express my sincere gratitude to Prof. Fosca Giannotti and Dr. Roberto Pellungrini whose support and contribution have been instrumental to the completion of this work. Their guidance, insightful feedbacks, and continuous encouragment significantly enhanced the depth and quality of this research.

---

[20] https://cdn.openai.com/openai-preparedness-framework-beta.pdf